\documentclass[journal,twoside]{IEEEtran}
\usepackage{cite}
\usepackage{amsmath}
\newtheorem{definition}{Definition}
\newtheorem{theorem}{Theorem}

\newtheorem{lemma}{Lemma}

\newtheorem{corollary}{Corollary}

\newcommand{\be}{\begin{equation}}
\newcommand{\ee}{\end{equation}}
\newcommand{\bea}{\begin{eqnarray}}
\newcommand{\eea}{\end{eqnarray}}
\newcommand{\T}{{\cal{T}}}

\begin{document}
%
\title{Asymptotic Entanglement Manipulation\\ of Bipartite Pure States}
%
%
\author{{Garry~Bowen~and~Nilanjana~Datta}
\thanks{This work was supported by the EPSRC (Research Grant GR/S92816/01).}%
\thanks{G. Bowen is with the Centre for Quantum Computation, Department of Applied Mathematics and Theoretical Physics, University of Cambridge, Cambridge CB3 0WA, UK (e-mail: gab30@damtp.cam.ac.uk).}
\thanks{N. Datta is with the Statistical Laboratory, Department of Pure Mathematics and Mathematical Statistics, University of Cambridge, Cambridge CB3 0WB, UK (e-mail: N.Datta@statslab.cam.ac.uk).}%
}
\markboth{Submitted to IEEE Transactions on Information Theory}{Asymptotic Entanglement Manipulation of Bipartite Pure States}
%

\maketitle

\begin{abstract}
Entanglement of pure states of bipartite quantum systems has been shown to have a unique measure in terms of the von Neumann entropy of the reduced states of either of its subsystems.  The measure is established under entanglement manipulation of an asymptotically large number of copies of the states.  In this paper, different asymptotic measures of entanglement assigned to arbitrary sequences of bipartite pure states are shown to coincide only when the sequence is \textit{information stable}, in terms of the quantum spectral information rates of the sequence of subsystem states.  Additional bounds on the optimal rates of 
entanglement manipulation protocols in quantum information are also presented, including bounds given by generalizations of the coherent information and the relative entropy of entanglement.
\end{abstract}


\IEEEpeerreviewmaketitle

\section{Introduction}

\PARstart{E}{ntanglement} in quantum information theory is a resource that has no counterpart in classical information theory.  Consequently, 
entanglement based protocols such as quantum cryptography \cite{ekert91}, quantum dense coding \cite{bennett92}, and quantum teleportation \cite{bennett93}, are unique to the domain of quantum information theory.

As for any resource, it is useful to have a measure of entanglement for quantum states.  A vast literature exists on various measures of 
entanglement for both bipartite and multipartite quantum states (see e.g. \cite{plenio05} for a review).

As entanglement is a non-local quantum resource, one of its fundamental property is that it cannot be reliably increased under local operations and classical communication (abbreviated to LOCC).  Therefore, if one state can be transformed into another by LOCC, then the target state must necessarily have no more entanglement than the original state.  By defining the entanglement $E$ of a \textit{maximally entangled state} of rank $M$,
\begin{equation}
|\Psi^+_M \rangle = \frac{1}{\sqrt{M}} \sum_{i=1}^M |i^A\rangle |i^B\rangle
\end{equation}
in ${\cal{H}}_A \otimes {\cal{H}}_B$ as being $E(|\Psi^+_M \rangle\langle \Psi^+_M|) = \log M$, we gain a benchmark state against which to measure the entanglement of other states.  Note that the base to which the logarithm is taken, simply determines the units in which entanglement is measured\footnote{Throughout this paper, we choose the logarithm to base $e$. We could
equally well choose an arbitrary base for
the logarithm. This would simply scale the unit of information.}.

Upper and lower bounds on the entanglement of an arbitrary bipartite state $\rho$ may then be constructed by determining the minimal rank of a maximally entangled state that can be transformed into $\rho$ by LOCC, and similarly by 
determining the largest rank maximally entangled state that $\rho$ may be transformed into by LOCC. 
We refer to 
the transformation of one entangled state to another, via LOCC alone, 
as entanglement manipulation. For pure bipartite states a theorem of Nielsen \cite{nielsen99} gives a criterion under which a pure bipartite state may be transformed into another pure bipartite state by LOCC alone. This provides a useful tool in determining the entanglement of these states.

As well as establishing bounds on entanglement for individual states, 
states may be assigned an asymptotic measure of entanglement. This is done by considering the
entanglement manipulation of the state $\varrho_n$ given by $n$ copies of the original bipartite state $\rho$, i.e.,
\begin{equation}
\varrho_n = \rho^{\otimes n}, \label{eqn:product}
\end{equation}
the $n$ copies represented by an $n$-fold tensor product.  The asymptotic 
measure of the entanglement of $\rho$ is then given by
\begin{equation}
\lim_{n\rightarrow \infty} \frac{1}{n} E(\rho^{\otimes n}).
\end{equation}
By relaxing the condition that the transformation to (or from) a maximally entangled state be exact for finite $n$, but requiring that the fidelity of the transformation approaches one as $n\rightarrow \infty$, we obtain the two asymptotic measures of entanglement called the entanglement of distillation (or distillable entanglement) $E_D(\rho)$ \cite{bennett96}, and the entanglement cost $E_C(\rho)$ \cite{hayden00}, respectively.

In the case of pure bipartite states $\phi^{AB} \in {\cal{H}}_A \otimes {\cal{H}}_B$, it has been shown that the unique measure of entanglement is given by \cite{bennett96b}
\begin{equation}
E = S(\rho),
\end{equation}
where $S(\rho) = -\mathrm{Tr} \rho \log \rho$ is the von Neumann entropy of the reduced state $\rho = \mathrm{Tr}_B \phi^{AB}$.  The uniqueness arises from the fact that the distillable entanglement $E_D$ and entanglement cost $E_C$ of any bipartite state $\rho$ represent limits for any asymptotic bipartite entanglement measure \cite{horodecki00a}.  That is, for any other asymptotic entanglement measure $E$ we have
\begin{equation}
E_D \leq E \leq E_C
\end{equation}
for any bipartite state $\rho$.
Moreover, it is known that the transformation to and from a maximally entangled state may be achieved at this rate with vanishing amounts of classical communication \cite{lo99}.

The practical ability to transform entanglement from one form to another is useful for many applications in quantum information theory.  However, it is not always justified to assume that the entanglement resource available, consists
of states which are multiple copies (and hence tensor products) of
a given entangled state. More generally, an entanglement resource is 
characterized by an arbitrary sequence of bipartite states which are 
not necessarily of the tensor product form (\ref{eqn:product}).
In order to examine entanglement manipulation for such resources
it is possible to use the tools provided by the \textit{information spectrum} methodology.  The information spectrum, derived in classical information theory by Verdu \& Han \cite{verdu94,han}, has been extended into quantum information by Hayashi \& Nagaoka \cite{nagaoka02,hayashi03}, initially in terms of quantum hypothesis testing.  The power of the information spectrum approach comes from the lack of assumptions made about the sources, channels or resources involved.

For an arbitrary sequences of states $\rho=\{\rho_n\}_{n=1}^\infty$, two 
real-valued quantities $\underline{S}(\rho)$ and $\overline{S}(\rho)$ can 
be defined (see Section \ref{infspectrum} or \cite{bowen06}). These are 
referred
to as the {\em{inf-spectral entropy rate}} and the {\em{sup-spectral entropy 
rate}} 
of $\rho$, respectively.
In this paper, we show that for arbitrary sequences of bipartite pure states $\phi = \{ \phi^{AB}_n \}_{n=1}^{\infty}$ the asymptotic entanglement is given by a single measure only when the sequence of reduced states (of either subsystem) 
is \textit{information stable}.  By information stable, we mean that the inf-spectral entropy rate of the sequence $\rho = \{ \rho_n \}_{n=1}^{\infty} = \{ \mathrm{Tr}_B\phi^{AB}_n \}_{n=1}^{\infty}$ is equal to the sup-spectral entropy rate, that is $\underline{S}(\rho) = \overline{S}(\rho)$.  Information stability is also known as the strong converse property.  If $\underline{S}(\rho) < \overline{S}(\rho)$ then a separation exists between entanglement measures.
We show this by proving that the entanglement cost of the sequence $\phi$ is given by
$\overline{S}(\rho)$, whereas it is known that the distillable entanglement is 
given by $\underline{S}(\rho)$ \cite{hayashi03a}.

Moreover, for information stable sequences the asymptotic entanglement may be expressed in the form
\begin{equation}
E = \lim_{n\rightarrow \infty} \frac{1}{n} S(\rho_n),
\end{equation}
the von Neumann entropy rate of the reduced states of either subsystem.

In addition, we provide bounds on entanglement distillation rates for sequences of \textit{arbitrary} bipartite states.  The bounds include information spectrum generalizations of the coherent information bounds \cite{schumacher96a}, under local operations involving no communication and LOCC bounds for one-way or two-way classical communication.  Further to this, an information spectrum generalization of the relative entropy of entanglement $E_R$ \cite{vedral97} is shown to provide an asymptotic upper bound on the distillable entanglement under arbitrary LOCC protocols.

In Section \ref{prelims} we outline some basic mathematic preliminaries.  Section \ref{bounds} contains proofs of the generalizations of the coherent information and relative entropy of entanglement bounds.  Following this, Section \ref{concent} contains a review of the entanglement concentration result of Hayashi \cite{hayashi03a} as well as a new proof of the weak converse.  Section \ref{dilut} shows that the entanglement cost for sequences of bipartite pure states is given by the sup-spectral entropy rate of the sequence of states on either subsystem.  Finally, in Section \ref{discuss} we give a unified presentation of what the combined results achieve in terms of the asymptotic entanglement of sequences of bipartite pure states.

\section{Mathematical Preliminaries}
\label{prelims}

Let ${\cal B}({\cal H})$ denote the algebra of linear operators acting on a finite--dimensional Hilbert space ${\cal H}$ of dimension $d$. Quantum states are represented by density matrices $\rho$, i.e.~positive operators of unit trace in ${\cal B}({\cal H})$.  Bipartite quantum states are states on Hilbert spaces ${\cal H}_A \otimes {\cal H}_B$, with $A$ and $B$ denoting the two parties sharing the state.  Sequences of bipartite states on $AB$ are considered to exist on Hilbert spaces ${\cal H}_A^{(n)} \otimes {\cal H}_B^{(n)}$ for $n \in \{1,2,3,...\}$.  In this paper, the shorthand notation $\phi^{AB}_n := |\phi^{AB}_n\rangle \langle \phi^{AB}_n|$ is used to denote density matrices of pure states.

\subsection{Fidelity and Reliable Transformations}

Let ${\cal{T}}_n$ be a quantum operation used for the transformation of an initial bipartite state $\rho_n$ to a bipartite pure state $|\sigma_n\rangle$.  
For the entanglement manipulation processes considered in this paper, 
${\cal{T}}_n$ either consists of local operations (LO) alone or LO with
one-way or two-way classical communication.
We define the fidelity of any entanglement manipulation process in terms of the overlap between the output state ${\cal{T}}_n(\rho_n)$ and the target state $|\sigma_n\rangle$. Specifically,
\begin{align}
F_n &:= \langle \sigma_n |{\cal{T}}_n(\rho_n)|\sigma_n\rangle \nonumber \\
&= \mathrm{Tr}\big[ {\cal{T}}_n(\rho_n) \sigma_n \big]
\end{align}
which is the square of the usual fidelity measure \cite{nielsen}.
An entanglement manipulation process is said to be \textit{reliable} if the asymptotic fidelity $\mathcal{F} := \liminf_{n\rightarrow \infty} F_n = 1$.

\subsection{Entanglement Rates}

The concept of reliable entanglement manipulation is then used to define two important asymptotic entanglement measures.

\begin{definition}
A real-valued number $R$ is said to be an \textit{achievable} distillation rate if $\forall \epsilon > 0,\, \exists N$ such that $\forall n \geq N$ a transformation exists that takes $\rho_n \rightarrow |\Psi^+_{M_n}\rangle \langle \Psi^+_{M_n}|$ with fidelity $F_n \geq 1-\epsilon$ and $\frac{1}{n}\log M_n \geq R$.
\end{definition}

\begin{definition}
The \textit{distillable entanglement} is the supremum of all achievable distillation rates,
\begin{equation}
E_D = \sup R
\end{equation}
for the required class of transformations (local operations only, or local operations with one-way or two-way communication).
\end{definition}

\begin{definition}
A real-valued number $R^*$ is said to be an \textit{achievable} dilution rate if $\forall \epsilon > 0,\, \exists N$ such that $\forall n \geq N$ a transformation exists that takes $|\Psi^+_{M_n}\rangle \langle \Psi^+_{M_n}| \rightarrow \rho_n$ with fidelity $F_n \geq 1-\epsilon$ and $\frac{1}{n}\log M_n \leq R^*$.
\end{definition}

\begin{definition}
The \textit{entanglement cost} is the infimum of all achievable dilution rates,
\begin{equation}
E_C = \inf R^*
\end{equation}
for the required class of transformations.
\end{definition}

\subsection{Spectral Projections}

The quantum information spectrum approach requires the extensive use of
spectral operators.  For a self-adjoint operator $A$ written in its spectral
decomposition $A = \sum_i \lambda_i |i\rangle \langle i|$ we define the
positive spectral projection on $A$ as
\begin{equation}
\{ A \geq 0 \} = \sum_{\lambda_i \geq 0} |i\rangle \langle i|
\end{equation}
the projector onto the eigenspace of positive eigenvalues of $A$.
Corresponding definitions apply for the other spectral projections $\{ A < 0
\}$, $\{ A > 0 \}$ and $\{ A \leq 0 \}$.  For two operators $A$ and $B$, we can
then define $\{ A \geq B \}$ as $\{ A - B \geq 0 \}$, and similarly for the
other ordering relations.

\subsection{Several Important Lemmas}

The following key lemmas are used repeatedly in the paper. For their proofs
see \cite{bowen06}.

\begin{lemma}
\label{lemma}
For self-adjoint operators $A$, $B$ and any positive operator $0 \leq P \leq I$
the inequality
\begin{equation}
\mathrm{Tr}\big[ P(A-B)\big] \leq \mathrm{Tr}\big[ \big\{ A \geq B \big\}
(A-B)\big]
\label{eqn:first_ineq}
\end{equation}
holds.
\end{lemma}

\begin{lemma}
\label{lemma2}
For self-adjoint operators $A$ and $B$, and any completely positive
trace-preserving (CPTP) map $\mathcal{T}$ the inequality
\begin{equation}
\mathrm{Tr}\big[ \{\mathcal{T}(A) \geq \mathcal{T}(B) \}\mathcal{T}(A-B)\big]
\leq \mathrm{Tr}\big[ \big\{ A \geq B \big\} (A-B)\big]
\label{eqn:second_ineq}
\end{equation}
holds.
\end{lemma}

\begin{lemma}
\label{cor0}
Given a state $\rho_n$ and a self-adjoint
operator $\omega_n$, we have
\be
\mathrm{Tr}\big[\{\rho_n \ge e^{n\gamma}\omega_n \} \omega_n \bigr]
\le e^{-n\gamma}.
\ee
for any real $\gamma$.
\end{lemma}

\subsection{Quantum Spectral Information Rates}
\label{infspectrum}

In the Quantum Information Spectrum approach, one defines spectral divergence
rates, which can be viewed as generalizations of the quantum relative entropy.
The spectral generalizations of the von Neumann entropy, the conditional
entropy and the mutual information can all be expressed as spectral 
divergence rates.
\begin{definition}
Given two sequences of states $\rho=\{\rho_n\}_{n=1}^\infty$ and
$\omega=\{\omega_n\}_{n=1}^\infty$, 
the quantum spectral sup-(inf-)divergence rates are defined in terms
of the difference operators $\Pi_n(\gamma) = \rho_n - e^{n\gamma}\omega_n$,  
as
\begin{align}
\overline{D}(\rho \| \omega) &= \inf \Big\{ \gamma : \limsup_{n\rightarrow \infty} \mathrm{Tr}\big[ \{ \Pi_n(\gamma) \geq 0 \} \Pi_n(\gamma) \big] = 0 \Big\} \label{supdiv} \\
\underline{D}(\rho \| \omega) &= \sup \Big\{ \gamma : \liminf_{n\rightarrow \infty} \mathrm{Tr}\big[ \{ \Pi_n(\gamma) \geq 0 \} 
\Pi_n(\gamma) \big] = 1 \Big\}, \label{infdiv}
\end{align}
respectively.
\end{definition}

The spectral entropy rates and the conditional spectral entropy rates
can be expressed as divergence rates with appropriate
substitutions for the sequence of operators $\omega = \{ \omega_n
\}_{n=1}^{\infty}$.  These are
\begin{align}
\overline{S}(\rho) &= -\underline{D}(\rho| I) \label{supent} \\
\underline{S}(\rho) &= -\overline{D}(\rho| I) 
\end{align}
and for sequences of bipartite state $\rho^{AB} = \{\rho^{AB}_n\}_{n=1}^\infty$,
\begin{align}
\overline{S}(A|B) &= -\underline{D}(\rho^{AB}| I^{A}\otimes \rho^B) \\
\underline{S}(A|B) &= -\overline{D}(\rho^{AB}| I^{A}\otimes \rho^B) \; .
\end{align}
In the above,
$I^{A}=\{I^A_n\}_{n=1}^\infty$ and $\rho^{A}=\{\rho^A_n\}_{n=1}^\infty$,
with $I^A_n$ being the identity operator in ${\cal{B}}({\cal{H}}_A^{(n)})$
and $\rho^A_n = \mathrm{Tr}_B \rho^{AB}_n$, the partial trace
being taken on the Hilbert space ${\cal{H}}_B^{(n)}$. Various properties and
relationships of these quantities are explored in \cite{bowen06}.

\section{Bounds on Entanglement}
\label{bounds}

For sequences of bipartite states we may obtain several bounds on the asymptotic entanglement.  The first family of bounds are generalizations of the coherent information bounds \cite{schumacher96a}.  The four inequalities in this family, implicitly contained in (\ref{coh_bound}), represent upper bounds on the distillable entanglement in the following cases respectively: local operations with $(i)$ no classical communication, $(ii)$ 
forward classical communication, $(iii)$ backward classical communication and
$(iv)$ two-way classical communication.
\begin{theorem}
The distillable entanglement for a sequence of bipartite states 
$\rho^{AB}=\{\rho^{AB}_n\}_{n=1}^\infty$,
is bounded above by
\begin{equation}
E_D (\rho^{AB}) \leq \max_{\mathcal{T}} I_B(\mathcal{T}(\rho^{AB}))
\label{coh_bound}
\end{equation}
where $I_B(\omega^{AB}) = -\overline{S}(A|B)$ the negative of the sup-conditional spectral entropy rate of the sequence $\omega^{AB}_n$, and $\mathcal{T} = \{{\mathcal{T}}_n\}_{n=1}^\infty$ is a sequence of maps representing either local operations on $A$, or, local operations with forward, backward, or two-way communication.
\end{theorem}

\begin{proof}
For any quantum operation $\mathcal{T}_n$, define $\omega_n^{AB} = \mathcal{T}_n(\rho_n^{AB})$. Then
\begin{align}
F_n &= \mathrm{Tr}\big[ |\Psi^+_{M_n}\rangle \langle \Psi^+_{M_n}|\mathcal{T}_n(\rho^{AB}_n) \big] \nonumber \\
&\leq \mathrm{Tr} \big[ \big\{ \omega^{AB}_n \geq e^{n\gamma}I^A_n\otimes\omega^{B}_n \big\} (\omega^{AB}_n - e^{n\gamma}I^A_n\otimes\omega^{B}_n) \big] \nonumber \\
&\phantom{=}\;+ e^{n\gamma} \mathrm{Tr}\big[|\Psi^+_{M_n}\rangle \langle \Psi^+_{M_n}|(I^A_n\otimes\omega^{B}_n)\big] \label{uselem1}\\
&\leq \mathrm{Tr} \big[ \big\{ \omega^{AB}_n \geq e^{n\gamma}I^A_n\otimes\omega^{B}_n \big\} (\omega^{AB}_n - e^{n\gamma}I^A_n\otimes\omega^{B}_n) \big] 
\nonumber\\
& \quad + \frac{e^{n\gamma}}{M_n}.
\end{align}
The first inequality follows from Lemma \ref{lemma} and the
second one is obtained by explicit evaluation of the last term in
(\ref{uselem1}). Substituting $\frac{1}{n}\log M_n \geq R = -\overline{S}(A|B) + \delta$ and $\gamma = -\overline{S}(A|B) + \delta/2$, the asymptotic fidelity is bounded above by $1-\epsilon_0$ for some $\epsilon_0 > 0$.
\end{proof}

The next theorem expresses a generalization of the bound on distillable entanglement given by the relative entropy of entanglement \cite{vedral98}.  The bound is not tight in general, although it reduces to the von Neumann entropy in the case of tensor products of pure states.
\begin{theorem}
\label{theorem1}
The two-way distillable entanglement is bounded above by the inf-spectral relative entropy of entanglement. Specifically,
\begin{equation}
E_D(\rho^{AB}) \leq \min_{\sigma \in \mathcal{D}} \underline{D}(\rho^{AB} \| \sigma^{AB} )
\end{equation}
where $\mathcal{D}$ is the set of sequences of states that are separable on $AB$.
\end{theorem}

\begin{proof}
For a maximally entangled state $|\Psi^+_{M_n}\rangle \langle \Psi^+_{M_n}|$ of rank $M_n$, the fidelity under two-way LOCC maps ${\cal{T}}_n$ is bounded by
\begin{align}
F_n &= \mathrm{Tr}\big[ |\Psi^+_{M_n}\rangle \langle \Psi^+_{M_n}|\mathcal{T}_n(\rho^{AB}_n) \big] \nonumber \\
&\leq \mathrm{Tr} \big[ \big\{ \mathcal{T}_n(\rho^{AB}_n) \geq \mathcal{T}_n(\omega^{AB}_n) \big\} \mathcal{T}_n(\rho^{AB}_n - \omega^{AB}_n) \big] \nonumber \\
&\phantom{=}\;+ \mathrm{Tr}\big[|\Psi^+_{M_n}\rangle \langle \Psi^+_{M_n}|\mathcal{T}_n(\omega^{AB}_n)\big] \\
&\leq \mathrm{Tr} \big[ \big\{ \rho^{AB}_n \geq \omega^{AB}_n \big\} \big(\rho^{AB}_n - \omega^{AB}_n\big) \big] \nonumber \\
&\phantom{=}\;+ \mathrm{Tr}\big[|\Psi^+_{M_n}\rangle \langle \Psi^+_{M_n}|\mathcal{T}_n(\omega^{AB}_n)\big].
\label{fid2}
\end{align}
The first term on the RHS of (\ref{fid2}) is obtained by using Lemma \ref{lemma2}. By choosing $\omega^{AB}_n = e^{n\gamma}\sigma^{AB}_n$, for any separable state $\sigma^{AB}_n$, the last term on RHS of (\ref{fid2}) is bounded by
\begin{equation}
\mathrm{Tr}\big[|\Psi^+_{M_n}\rangle \langle \Psi^+_{M_n}|\mathcal{T}_n(\omega^{AB})\big] \leq \frac{e^{n\gamma}}{M_n},
\end{equation}
and hence
\begin{align}
F_n &\leq \mathrm{Tr} \big[ \big\{ \rho^{AB}_n \geq e^{n\gamma}\sigma^{AB}_n \big\} \big(\rho^{AB}_n - e^{n\gamma}\sigma^{AB}_n\big) \big] + \frac{e^{n\gamma}}{M_n}
\end{align}
for all separable states $\sigma^{AB}_n$. Choose a sequence of states $\sigma^{AB}= \{\sigma^{AB}_n\}_{n=1}^\infty$, such that the divergence rate $\underline{D}(\rho^{AB} \| \sigma^{AB} )$ is minimized. Then choosing $\gamma = \underline{D}(\rho^{AB} \| \sigma^{AB} ) + \delta/2$, for some $\delta >0$, 
we can see that for any rate $\frac{1}{n}\log M_n \geq R = \underline{D}(\rho^{AB} \| \sigma^{AB} ) + \delta$, the asymptotic fidelity is bounded above by $1-\epsilon_0$ for some $\epsilon_0 > 0$.
\end{proof}

\section{Entanglement Concentration}
\label{concent}

Entanglement concentration is the protocol in which two parties, Alice and 
Bob (say), share
a sequence of partially entangled pure states 
$\{|\phi_n\rangle\}_{n=1}^\infty$, with $|\phi_n\rangle \in
{\cal{H}}_A^{(n)} \otimes {\cal{H}}_B^{(n)}$
and they wish to convert them into a sequence of maximally entangled pure
states $\{|\Psi_{M_n}^+\rangle\}_{n=1}^\infty$, where
$$|\Psi_{M_n}^+\rangle := \frac{1}{\sqrt{M_n}} \sum_{i=1}^{M_n}
| i_n^A\rangle \otimes |i_n^B\rangle
\, \in {\cal{H}}_A^{(n)} \otimes {\cal{H}}_B^{(n)},$$
by LOCC alone.

Entanglement concentration may be utilized to determine the distillable entanglement of sequences of pure bipartite states.  The main result presented in this section is the following theorem \cite{hayashi03a}.
\begin{theorem}[Hayashi]
\label{conc_theorem}
The distillable entanglement of a sequence of bipartite pure states $\phi^{AB} = \{ \phi^{AB}_n \}_{n=1}^{\infty}$, is given by
\begin{equation}
E_D = \underline{S}(\rho)
\end{equation}
for $\rho = \{ \mathrm{Tr}_A \phi^{AB}_n \}_{n=1}^{\infty}$, the sequence of subsystem states.
\end{theorem}

The proof of Theorem \ref{conc_theorem} requires the following three lemmas.
\begin{lemma} {\em{(Coding)}} 
Given a sequence of bipartite pure states $\phi^{AB} = \{ \phi^{AB}_n \}_{n=1}^{\infty}$, for any $\epsilon >0$ and $\delta > 0$, there exists an $N$ such that $\forall n \geq N$, maximally entangled states may be generated at a rate
\begin{equation}
R \geq \underline{S}(\rho) - \delta,
\end{equation}
with probability of failure
\begin{equation}
P_{\mathrm{fail}} \leq \epsilon.
\end{equation}
Here $\rho = \{ \rho_n \}_{n=1}^{\infty}$ is the sequence of reduced states of the pure bipartite states in $\phi^{AB}$.
\end{lemma}
\begin{proof}
Let the bipartite state $|\phi_n\rangle \in {\cal{H}}_A^{(n)} \otimes
{\cal{H}}_B^{(n)}$ have a spectral decomposition
\begin{equation}
|\phi_n\rangle = \sum_i \sqrt{\lambda_{n,i}} |i^A_n\rangle \otimes 
|i^B_n\rangle. 
\end{equation}
Define projection operators $Q_n = \{ \rho_n < e^{-n\gamma}I_n \}$ and $\overline{Q}_n = I_n - Q_n$. The first step of the protocol is for one of the parties 
(say, Alice) to do a von Neumann measurement, described by the 
projection operators $Q_n$ and $\overline{Q}_n$, on her part of the shared
bipartite state $|\phi_n\rangle$. If the outcome of the measurement corresponds to $\overline{Q}_n$, then the protocol is aborted as unsuccessful.  This occurs with probability $P_{\mathrm{fail}} = \mathrm{Tr}[\overline{Q}_n\rho_n]$.

If the outcome of the measurement corresponds to $Q_n$, then
the post-measurement state is given by
\begin{equation}
|\psi_n\rangle:= \frac{1}{\sqrt{\mathrm{Tr}[Q_n\rho_n]}} \sum_{\lambda_{n,i} < e^{-n\gamma}} \sqrt{\lambda_{n,i}} |i_n^A\rangle |i_n^B\rangle,
\end{equation}
and each of the eigenvalues of the reduced density matrix of this state is bounded above by
\begin{equation}
\lambda_{n,i} \leq \frac{e^{-n\gamma}}{\mathrm{Tr}[Q_n\rho_n]}
\label{eig_bound}
\end{equation}

Nielsen's majorization theorem \cite{nielsen99} states that a bipartite pure state $\Phi$ with subsystem state $\rho$, may be converted by (one-way) LOCC into the pure state $\Psi$ with subsystem state $\omega$, if and only if the ordered eigenvalues of $\rho$ are \textit{majorized} by those of $\omega$.  Specifically,
\begin{equation}
\sum_{i=1}^k \lambda_{\rho}^k \leq \sum_{i=1}^k \lambda_{\omega}^k
\end{equation}
for all $k$, where $\lambda_{\rho}^j \geq \lambda_{\rho}^{j+1}$ and similarly $\lambda_{\omega}^j \geq \lambda_{\omega}^{j+1}$, for all $j \in \{1,..., d-1\}$.

It follows from Nielsen's theorem that the state $|\psi_n\rangle$ may be transformed by one-way LOCC into the maximally entangled state $|\Psi_{M_n}^+\rangle$ of rank
\begin{equation}
M_n = \lfloor \mathrm{Tr}[Q_n\rho_n] e^{n\gamma} \rfloor
\end{equation}
as the eigenvalues all obey the inequality in (\ref{eig_bound}).
This concludes the protocol.

For any choice of $\gamma$ such that $P_{\mathrm{fail}} = \mathrm{Tr}[\overline{Q}_n\rho_n] < 1/2$, and large enough $n$, the achievable rate is given by
\begin{align}
R &= \frac{1}{n} \log M_n \nonumber \\
&\geq \gamma - \frac{2}{n}\big(\epsilon_n + e^{-n\gamma}\big)
\end{align}
where $\epsilon_n = 1 - \mathrm{Tr}[Q_n\rho_n]$. This can be seen by using the fact that $\lfloor C e^{n\gamma} \rfloor \geq C e^{n\gamma} - 1 = e^{n\gamma}(C - e^{-n\gamma})$ and $\log (1-x) \geq -2x$ for $x \leq 3/4$.

Choosing $\gamma = \underline{S}(\rho) - \delta/2$ implies that $\mathrm{Tr}[Q_n\rho_n] \rightarrow 1$ as $n\rightarrow \infty$, and therefore we can choose an $N$ such that both
\begin{equation}
\mathrm{Tr}[Q_n\rho_n] \geq 1 - \epsilon
\end{equation}
and
\begin{equation}
\frac{1}{n}(\epsilon_n + e^{-n\gamma}) \leq \frac{\delta}{4}
\end{equation}
whenever $n \geq N$.
\end{proof}

To prove the weak converse for entanglement concentration we require the following property of the conditional spectral entropy for bipartite pure states.
\begin{lemma}
\label{lemma4}
Let $\sigma_n^{AB}:= \T_n^{AB}\big(\phi_n^{AB}\big)$,
where $\phi_n^{AB}\in {\cal{B}}({\cal{H}}_A^{(n)} \otimes{\cal{H}}_B^{(n)})$
and $\T_n^{AB}$ denotes any LOCC operation. Let ${\overline{S}}_\phi (A|B)$
denote the sup-spectral conditional entropy ${\overline{S}}(A|B)$ for the
sequence of pure state $\phi^{AB}:=\{\phi_n^{AB} \}_{n=1}^\infty$. Let 
${\overline{S}}_\sigma (A|B)$ be the corresponding quantity for the 
sequence $\sigma^{AB}:=\{\sigma_n^{AB}\}_{n=1}^\infty$. Then the inequality
\begin{equation}
{\overline{S}}_\phi (A|B) \le {\overline{S}}_\sigma (A|B).
\end{equation}
holds.
\end{lemma}
\begin{proof}
Note that $$-{\overline{S}}_\phi (A|B) := \underline{D}\big(\phi^{AB}| I^A
\otimes \mathrm{Tr}_A \phi^{AB} \big),$$  
and 
\begin{align}
-{\overline{S}}_\sigma (A|B) &:= \underline{D}\big(\sigma^{AB}| I^A
\otimes \mathrm{Tr}_A \sigma^{AB} \big)\nonumber\\
&= \underline{D}\big(\T^{AB}(\phi^{AB})| I^A
\otimes \mathrm{Tr}_A \T^{AB}(\phi^{AB}) \big),\nonumber\\
\end{align}
where $\T^{AB}:= \{\T_n^{AB}\}_{n=1}^\infty$, a sequence of LOCC operations 
and $\mathrm{Tr}_A \T^{AB}(\phi^{AB})
= \{\mathrm{Tr}_{A_n} \T_n^{AB}(\phi^{AB}_n)\}_{n=1}^\infty$, with
$\mathrm{Tr}_{A_n}$ denoting the partial trace over the 
Hilbert space ${\cal{H}}_A^{(n)}$.
The action of the LOCC operation $\T_n^{AB}$ on the state $\phi^{AB}_n$
can be expressed as follows (see \cite{lo01}):
\begin{equation}
\T_n^{AB}(\phi^{AB}_n)= \sum_j (U_{n,j} \otimes K_{n,j})(\phi^{AB}_n)
(U_{n,j}^\dagger \otimes K_{n,j}^\dagger),
\label{LOCC}
\end{equation}
where the $U_{n,j}$ are unitary operators and $K_{n,j}$ are operators such that $\sum_j K_{n,j}^\dagger K_{n,j} = I$.  Denoting the reduced state $\omega_n^B = \mathrm{Tr}_A\phi_n^{AB}$, it then follows that
\begin{align}
\T_n^{AB}\big(I^A_n \otimes \omega_n^B \big) &= \sum_j U_{n,j}U_{n,j}^\dagger \otimes K_{n,j} \omega^{B}_n K_{n,j}^\dagger \nonumber \\
&= I^A_n \otimes \sum_j K_{n,j} \omega^{B}_n K_{n,j}^\dagger \nonumber \\
&= I^A_n \otimes \sigma^B_n
\end{align}
and hence,
\begin{align}
-{\overline{S}}_\sigma (A|B) &= \underline{D}\big(\sigma^{AB}| I^A \otimes \mathrm{Tr}_A \sigma^{AB} \big) \nonumber \\
&= \underline{D}\big(\T^{AB}(\phi^{AB})| \T^{AB}(I^A \otimes \mathrm{Tr}_A \phi^{AB}) \big) \nonumber \\
&\leq \underline{D}\big(\phi^{AB} | I^A \otimes \mathrm{Tr}_A \phi^{AB} \big) \nonumber \\
&= -{\overline{S}}_\phi (A|B)
\end{align}
where the inequality follows from Lemma \ref{lemma2}.
\end{proof}

It is then straightforward to show the weak converse.
\begin{lemma} {\em{(Weak Converse)}}
Any entanglement concentration protocol with rate $R > \underline{S}(\rho)$
is not achievable. Here $\rho = \{ \mathrm{Tr}_A \phi^{AB}_n \}_{n=1}^{\infty}$, with $\phi^{AB}_n$ denoting 
the pure bipartite initial states of the entanglement concentration process.
\end{lemma}
\begin{proof}
Combining Theorem \ref{theorem1} with the Chain Rule \cite{bowen06}
\begin{align}
{\underline{S}}(AB)- {\underline{S}}(B) &\leq {\overline{S}}(A|B)\nonumber\\
&\leq {\overline{S}}(AB) - {\underline{S}}(B),
\end{align}
and the fact that ${\underline{S}}(AB)= 0 = {\overline{S}}(A|B)$ for 
sequences of pure states on $AB$, yields the identity ${\overline{S}}(A|B)
= - {\underline{S}}(B)$. This along with 
Lemma \ref{lemma4} implies that for any rate
\begin{equation}
R > -\overline{S}_{\phi}(A|B) = \underline{S}(\rho),
\end{equation}
where $\rho = \{ \mathrm{Tr}_A \phi^{AB}_n \}_{n=1}^{\infty}$, 
the asymptotic fidelity is bounded above by $1-\epsilon_0$ for some $\epsilon_0 > 0$. 
\end{proof}

The strong converse rate for entanglement concentration is defined as the infimum of all rates, such that any distillation protocol of that rate has asymptotic fidelity $\mathcal{F} = 0$.
\begin{corollary}
The strong converse rate for entanglement concentration is given by
\begin{equation}
E^*_D = \overline{S}(\rho)
\end{equation}
for $\rho$ the sequence of subsystem states.
\end{corollary}
\begin{proof}
The proof follows the proof of the coding and weak converse.
\end{proof}

\section{Entanglement Dilution}
\label{dilut}

Entanglement dilution is the protocol which is essentially opposite to entanglement
concentration.  Here the two parties, Alice and Bob, share
a sequence of maximally entangled states
$\{|\Psi_{M_n}^+\rangle\}_{n=1}^\infty$, where
$$|\Psi_{M_n}^+\rangle := \frac{1}{\sqrt{M_n}} \sum_{i=1}^{M_n}
| i\rangle \otimes |i\rangle
\, \in {\cal{H}}^{(n)}_A \otimes {\cal{H}}_B^{{(n)}},$$
and wish to convert them into a sequence of non-maximally entangled pure
states $\{|\phi_n\rangle\}_{n=1}^\infty$, with $|\phi_n\rangle \in
{\cal{H}}^{(n)}_A \otimes {\cal{H}}_B^{{(n)}}$, and
with corresponding reduced density matrices $\rho_n^A \in {\cal{B}}(
 {\cal{H}}_A^{(n)})$.

Let the bipartite state $|\phi_n\rangle$ have $N_n$ non-zero
eigenvalues and let its spectral decomposition be given by
\begin{equation}
| \phi_n\rangle = \sum_{i=1}^{N_n} \sqrt{\lambda_{n,k}} |k\rangle \otimes |k\rangle,
\end{equation}
where the eigenvalues $\lambda_{n,k}$ are arranged in descending order, i.e.,
$\lambda_{n,1} \ge \lambda_{n,2} \ge \ldots \ge \lambda_{n,{N_n}}$. If $M_n \ge N_n$,
then Alice can perfectly teleport the state $|\phi_n\rangle$ to Bob,
using her part of the maximally entangled state $|\Psi^+_{M_n}\rangle$.
Hence, in this case the fidelity of the entanglement dilution
protocol is equal to unity.

However, if $M_n < N_n$, then Alice can perfectly teleport only the unnormalized
truncated state
\begin{equation}
| \Phi_{M_n} \rangle := P_{M_n}|\phi_n\rangle =  \sum_{i=1}^{M_n}
\sqrt{\lambda_{n,k}} |k\rangle \otimes |k\rangle.
\end{equation}
Here $P_{M_n}$ denotes the orthogonal projection onto the
${M_n}$ largest eigenvalues of $|\phi_n\rangle$. In this case the fidelity
is bounded below by
\begin{align}
F_n &\geq  \big|\langle \phi_n|\Phi_{M_n}\rangle \big|^2\nonumber\\
&= \Big| \sum_{k=1}^{N_n}  \sum_{j=1}^{M_n} \sqrt{\lambda_{n,k}}\sqrt{\lambda_{n,j}}
\langle k|j\rangle \langle k|j\rangle \Big|^2
\nonumber\\
&= \Big| \sum_{k=1}^{M_n}\lambda_{n,k} \Big|^2.
\end{align}
Although this protocol initially appears to be far from optimal, we show, in the proof of the following theorem, that for asymptotically reliable entanglement 
dilution, it is sufficient to obtain rates arbitrarily close to the entanglement cost.

\begin{theorem}
\label{cost_theorem}
The entanglement cost of a sequence of pure bipartite target states $\phi^{AB} = \{ \phi^{AB}_n \}_{n=1}^{\infty}$, is given by
\begin{equation}
E_C = \overline{S}(\rho),
\end{equation}
where $\rho = \{ \mathrm{Tr}_A \phi^{AB}_n \}_{n=1}^{\infty}$, the sequence of subsystem states.
\end{theorem}

The proof is contained in the following two lemmas.

\begin{lemma} {\em{(Coding)}}
For any $\delta > 0$, the dilution rate $R^* = \overline{S}(\rho)+\delta$ is achievable.
\end{lemma}
\begin{proof}
Suppose Alice
teleports the unnormalized states
\begin{equation}
| {\Phi_{n}} \rangle := Q_n|\phi_n\rangle,
\end{equation}
where $Q_n := \{\rho_n^A \ge e^{-n\alpha} I^A_n \} \otimes I_n^B$, and $\alpha$ is
a real
number satisfying $\alpha > \overline{S}(\rho^A)$, with $\rho^A$ being
the sequence of reduced density matrices of $|\phi_{n} \rangle\langle
\phi_{n}|$, i.e., $\rho^A = \{\rho_n^A\}_{n=1}^\infty$. From the
definition of $\overline{S}(\rho^A)$ it follows that
\begin{equation}
\lim_{n\rightarrow \infty} \mathrm{Tr}\big[ \{ \rho_n^A \geq e^{-n\alpha}I_n^A
\}
\rho_n^A \big] = 1.
\end{equation}
Hence, in this case the fidelity is given by
\begin{align}
F_n &\geq \langle \phi_n | Q_n |\phi_n\rangle \nonumber\\
&= \mathrm{Tr}\big[ \big(\{\rho_n^A \ge e^{-n\alpha }I_n^A\} \otimes I_n^B
\big) |\phi_{n} \rangle\langle \phi_{n}|\big] \nonumber\\
&= \mathrm{Tr}\big[ \{ \rho_n^A \geq e^{-n\alpha}I_n^A
\} \rho_n^A \big] \rightarrow 1 \,\, {\hbox{as }}\, n \rightarrow \infty.
\end{align}

Since $\mathrm{Tr}[ \{ \rho_n^A \geq e^{-n\alpha}I_n^A \} ] \leq
e^{n\alpha}$, then for large enough $n$ we can choose $e^{n\overline{S}(\rho)} < M_n \leq e^{n(\overline{S}(\rho)+\delta)}$ and entanglement
dilution with a sequence of maximally entangled states
$\{|\Psi_{M_n}^+\rangle\}_{n=1}^\infty$ of rank
$M_n$, is asymptotically achievable.
\end{proof}

\begin{lemma} {\em{(Weak Converse)}}
Any entanglement dilution protocol with a rate $R^* < \overline{S}(\rho)$ is not reliable.
\end{lemma}
\begin{proof}
Let ${\cal{T}}^{AB}_n$ denote any LOCC operation used for transforming
the maximally entangled state $|\Psi_{M_n}^+\rangle \in {\cal{H}}_A^{(n)} 
\otimes
{\cal{H}}_B^{(n)}$ to a partially
entangled state $|\phi_n\rangle$ in this Hilbert space.
Using (\ref{LOCC}), the fidelity of this transformation is expressible as
\begin{align}
F_n &= \mathrm{Tr}\big[|\phi_n\rangle\langle\phi_n| {\cal{T}}^{AB}_n\big(|\Psi_{M_n}^+\rangle \langle\Psi_{M_n}^+| \big)\big]\nonumber\\
&= \langle\phi_n|\sum_k(K_{n,k} \otimes U_{n,k})|\Psi_{M_n}^+\rangle \langle\Psi_{M_n}^+|
(K_{n,k} \otimes U_{n,k})^\dagger|\phi_n\rangle\nonumber\\
&= \sum_k \big| \langle\phi_n|(K_{n,k} \otimes U_{n,k})|\Psi_{M_n}^+\rangle \big|^2.
\end{align}

Let the state $|\phi_n\rangle$ have a Schmidt decomposition 
$|\phi_n\rangle = \sum_i \sqrt{\lambda_{n,i}} 
|i\rangle\otimes |i\rangle$.  Then
\begin{align}
 U^B_{n,k}|\Psi_{M_n}^+\rangle &= \frac{1}{\sqrt{{M_n}}} \sum_{\widetilde j=1}^{M_n} |\widetilde{j}^A\rangle
 U^B_{n,k}|\widetilde{j}^B\rangle
\nonumber\\
&=\frac{1}{\sqrt{{M_n}}} P^A_{M_n} \sum_{j=1}^N W^A|{j}^A\rangle
 U^B_{n,k} W^B|{j}^B\rangle,\nonumber\\
&= \frac{1}{\sqrt{{M_n}}}P^A_{M_n} \sum_{j=1}^N V^A_{n,k} |{j}^A\rangle|{j}^B\rangle,
\end{align}
where $N = \dim \mathcal{H}^{(n)}$, $W|{j}\rangle= |\widetilde{j}\rangle$, $V_{n,k} = (U_{n,k} W)^T W$
and $P^A_{M_n} = \sum_{\widetilde j=1}^{M_n}|\widetilde{j}^A\rangle 
\langle {\widetilde j}^A|$.
Here we have used the relation
$$\sum_j |j\rangle \otimes U|j\rangle = U^T |j\rangle \otimes |j\rangle$$
for $U$ unitary and $\{ |j\rangle \}$ an orthonormal basis.
Hence,
\begin{align}
\langle \phi_n| K_{n,k} \otimes U_{n,k} |\Psi_{M_n}^\dagger\rangle
&= \sum_{ij} \frac{{\sqrt{\lambda_{n,i}}}}{{\sqrt{{M_n}}}}
\langle i |K_{n,k} P_{M_n} V_{n,k} |j\rangle \langle i |j \rangle\nonumber\\
&= \sum_{i} \sqrt{ \frac{{\lambda_{n,i}}}{{M_n}}} 
\langle i |K_{n,k} P_{M_n} V_{n,k} |i\rangle\nonumber\\
&=  \mathrm{Tr}\Bigl[ \frac{1}{\sqrt{{M_n}}} {\sqrt{\sigma_n}}K_{n,k} P_{M_n} V_{n,k}  \Bigr],
\end{align}
where $\sigma_n= \mathrm{Tr}_B  |\phi_n\rangle\langle \phi_n | = 
\sum_i \lambda_{n,i} |i\rangle \langle i|$.
Using the Cauchy Schwarz inequality we have
\begin{align}
F_n&= \sum_k \Big| \mathrm{Tr}\bigl[ \frac{1}{\sqrt{{M_n}}}{\sqrt{\sigma_n}}K_{n,k} P_{M_n} V_{n,k}
\bigr]\Big|^2\nonumber\\
&= \sum_k \Big| \mathrm{Tr}\bigl[ \frac{1}{\sqrt{{M_n}}}K_{n,k} P_{M_n}\cdot P_{M_n} V_{n,k} {\sqrt{\sigma_n}}
\bigr]\Big|^2\nonumber\\
&\leq \sum_k  \frac{1}{{{M_n}}}\mathrm{Tr}\bigl[P_{M_n} K_{n,k}^\dagger K_{n,k}\bigr] \cdot
\mathrm{Tr}\bigl[\sigma_n V_{n,k}^\dagger P_{M_n} V_{n,k}\bigr]\nonumber\\
&\leq \frac{1}{{{M_n}}}\mathrm{Tr}[P_{M_n}] \cdot \max_{k'} \mathrm{Tr}\bigl[
  V_{n,k'}^\dagger P_{M_n} V_{n,k'}\sigma_n\bigr]
\nonumber\\
&= \frac{1}{{{M_n}}}\mathrm{Tr}[P_{M_n}]  \cdot \max_{k'} 
\mathrm{Tr}\bigl[P_{n,k'}{\sigma_n}\bigr]\nonumber\\
&= \max_{k'} 
\mathrm{Tr}\bigl[P_{n,k'}{\sigma_n}\bigr],
\end{align}
since $\mathrm{Tr}[P_{M_n}] ={M_n}$. In the above, $P_{n,k'}=V_{n,k'}^\dagger P_{M_n} V_{n,k'}$.

Using Lemma \ref{lemma}, we have
\begin{align}
\mathrm{Tr}\bigl[P_{n,k'}{\sigma_n}\bigr]&=  
\mathrm{Tr}\bigl[P_{n,k'}\bigl({\sigma_n}- e^{-n\gamma}I_n\bigr)\bigr]
+ e^{-n\gamma}\mathrm{Tr}P_{n,k'},\nonumber\\
&\leq \mathrm{Tr}\bigl[ \{\sigma_n \ge e^{-n\gamma}I_n\}(\sigma_n -
e^{-n\gamma}I_n)\bigr] \nonumber\\
& \quad + e^{-n\gamma}\mathrm{Tr}P_{n,k'}
\nonumber\\
&= \mathrm{Tr}\bigl[ \{\sigma_n \ge e^{-n\gamma}I_n\}(\sigma_n -
e^{-n\gamma}I_n)\bigr] \nonumber \\
&\quad + {M_n}e^{-n\gamma},
\end{align}
since $\mathrm{Tr}[P_{n,k'}]=\mathrm{Tr}[P_{{M_n}}]={M_n}$. Hence for ${M_n} \le e^{nR}$ we have
\begin{equation}
F_n \leq {\mathrm{Tr}}\bigl[\{\sigma_n \ge e^{-n\gamma}I_n\}(\sigma_n -
e^{-n\gamma}I_n)\bigr]
+ e^{-n(\gamma - R)}.
\label{last2}
\end{equation}
Choosing a number $\gamma$ and $\delta > 0$ such that $R = \gamma + \delta <
\overline{S}(\sigma)$, for $\sigma = \{\sigma_n\}_{n=1}^\infty$, the second term on RHS of 
(\ref{last2}) tends to zero as $n
\rightarrow \infty$. However, since $\gamma < {\overline{S}}(\sigma)$
the first term on RHS of (\ref{last2}) does not converge to $1$ as
$n \rightarrow \infty$. Hence, the asymptotic fidelity ${\cal{F}}$
is not equal to $1$.
\end{proof}

The strong converse rate for entanglement dilution is the supremum of all rates such that any dilution protocol has asymptotic fidelity $\mathcal{F} = 0$.
\begin{corollary}
The strong converse rate for entanglement dilution is given by
\begin{equation}
E^*_C = \underline{S}(\rho)
\end{equation}
for $\rho$ the sequence of subsystem states.
\end{corollary}
\begin{proof}
The proof follows as for the coding and weak converse.
\end{proof}

\section{Discussion}
\label{discuss}

As explicitly shown in Lemma 3 of \cite{bowen06a}, the von Neumann entropy rate is bounded above and below by the sup-spectral entropy and inf-spectral entropy rates, respectively.  For any sequence of bipartite pure states that is information stable on its subsystems, this implies
\begin{equation}
\underline{S}(\rho) = \lim_{n\rightarrow \infty} \frac{1}{n}S(\rho_n) = \overline{S}(\rho)
\end{equation}
and the asymptotic entanglement of the sequence is given by the von Neumann entropy rate of the subsystem states
\begin{equation}
E = \lim_{n\rightarrow \infty} \frac{1}{n}S(\rho_n) \; .
\label{elim}
\end{equation}
The set of information stable sequences includes all those 
sequences whose subsystem states are stationary and ergodic.
For the i.i.d. case, i.e., one in which $\phi_n^{AB}=  \phi^{\otimes n}$,
where $\phi$ is a pure state on ${\cal{H}}_A \otimes {\cal{H}}_B$, 
(\ref{elim}) reduces to
\begin{equation}
E = S(\rho),
\end{equation} for $\rho={\mathrm{Tr}}_B \phi$. 

An example of a sequence of pure bipartite states which is not information stable are those that have subsystem states that can be represented as mixtures of tensor product states with different von Neumann entropies, i.e., a sequence with subsystem states
\begin{equation}
\rho_n = t\, \sigma^{\otimes n} + (1-t)\, \omega^{\otimes n}
\end{equation}
with $t \in (0,1)$, such that $S(\sigma) < S(\omega)$.  From the results in Section III-B of \cite{bowen06a} it then follows that
\begin{equation}
E_D = S(\sigma) < S(\omega) = E_C
\end{equation}
for this sequence of states.

Furthermore, by examining the rates achievable for dense coding through a noiseless channel \cite{bowen06a}, it is easily seen that the capacity of the noiseless channel assisted by a sequence of shared bipartite pure states, is given by
\begin{align}
C_{DC} &= \log d - \overline{S}_{\phi}(A|B) \nonumber \\
&= \log d + \underline{S}(\rho) \nonumber \\
&= \log d + E_D \; ,
\end{align}
where $\rho = \{ \mathrm{Tr}_A \phi^{AB}_n \}_{n=1}^{\infty}$.
Hence, the dense coding capacity is enhanced over the capacity of the noiseless channel ($C = \log d$) by the distillable entanglement of the shared sequence of states.  In this regard, the distillable entanglement represents the usefulness of the shared states as an entanglement resource.


\end{document}